\documentclass[preprint2,twoside]{hwo}

\usepackage{graphicx}
\usepackage{amsmath}

\input{hwo.h}

\begin{document}

\title{\textbf{\LARGE Living Worlds Working Group: Surface Biosignatures on Potentially Habitable Exoplanets}}
% List all primary authors here. Contributing authors may be placed here
% or in a section below, at the discretion of the primary author.
% Please include the email address for the corresponding author.
\author {\textbf{\large Niki Parenteau,$^{1}$ Anna Grace Ulses,$^2$   Connor Metz,$^3$ Nancy Y. Kiang,$^4$ Ligia F. Coelho,$^5$ Edward Schwieterman,$^6$ Jonathan Grone,$^7$ Giulia Roccetti,$^8$ Svetlana Berdyugina,$^9$ Eleonora Alei,$^{10}$ Lucas Patty,$^7$ Emilie Lafleche,$^{11}$ Taro Matsuo,$^{12}$ Dawn Cardace,$^{13}$ Schuyler Borges,$^{14}$ Avi Mandel,$^{10}$ Kenneth Gordon,$^{15}$ Joshua Krissansen-Totton,$^2$ Giada Arney $^{10}$}}
\affil{$^1$\small\it Exobiology Branch $\vert$ NASA Ames Research Center, Moffett Field, CA, USA; \email{mary.n.parenteau@nasa.gov}}
\affil{$^2$\small\it University of Washington, Seattle, WA, USA.}
\affil{$^3$\small\it University of Michigan, Ann Arbor, MI, USA.}
\affil{$^4$\small\it NASA Goddard Institute for Space Studies, New York, NY, USA.}
\affil{$^5$\small\it Cornell University, Ithaca, NY, USA.}
\affil{$^6$\small\it University of California, Riverside, Riverside, CA, USA.}
\affil{$^7$\small\it University of Bern, Bern, Switzerland.}
\affil{$^8$\small\it Ludwig-Maximilian University of Munich, Munich, Germany.}
\affil{$^9$\small\it IRSOL, Universita Svizzera italiana, Locarno, Switzerland.}
\affil{$^{10}$\small\it Goddard Space Flight Center, Greenbelt, MD, USA.}
\affil{$^{11}$\small\it Purdue University, West Lafayette, IN, USA.}
\affil{$^{12}$\small\it Nagoya University, Nagoya, Japan.}
\affil{$^{13}$\small\it University of Rhode Island, Kingston, RI, USA.}
\affil{$^{14}$\small\it Northern Arizona University, Flagstaff, AZ, USA.}
\affil{$^{15}$\small\it University of Central Florida, Orlando, FL, USA.}

\author{\small{\bf Contributing Authors:} Bill Philpot (Cornell University), Jacob Lustig-Yeager (Johns Hopkins University Applied Physics Laboratory), Stephanie Olson (Purdue University), Marc Neveu (NASA Goddard Space Flight Center), José A. Caballero (Spanish Centro de Astrobiología [CSIC-INTA]), Yuka Fujii (National Astronomical Observatory of Japan), William Sparks (SETI Institute), Ludmila Carone (Space Research Institute, Austrian Academy of Sciences), Mariano Battistuzzi (University of Padova, Italy), Daniel Whitt (NASA Ames Research Center)}     

% Please add the names of endorsers in the format "Joseph Jensen (Utah Valley University), " separated by commas.
%\author{\footnotesize{\bf Endorsed by:}
%List endorsers in alphatical order here. Endorser 1 (affiliation), Endorser 2 (affilation), Endorser 3 (affiliation)
%}

% This section is for ADS Processing.  There must be one line per author. Leave them commented out for the present. They will be included later.
%\paperauthor{Sample~Author1}{Author1Email@email.edu}{ORCID_Or_Blank}{Author1 Institution}{Author1 Department}{City}{State/Province}{Postal Code}{Country}
%\paperauthor{Sample~Author2}{Author2Email@email.edu}{ORCID_Or_Blank}{Author2 Institution}{Author2 Department}{City}{State/Province}{Postal Code}{Country}
%\paperauthor{Sample~Author3}{Author3Email@email.edu}{ORCID_Or_Blank}{Author3 Institution}{Author3 Department}{City}{State/Province}{Postal Code}{Country}

% Please provide entries for the Author index; leave them commented out for now.
%\aindex{Pacucci, F.}

\begin{abstract}
 
Surface biosignatures are planetary-scale spectral features resulting from absorption and/or scattering of radiation by organisms containing photosynthetic and non-photosynthetic pigments. The canonical vegetation red edge (VRE) is so-called due to the sharp contrast in visible light absorbance by light harvesting pigments in plant leaves versus their high reflectance in the NIR. This secondary class of biosignatures can be used to corroborate atmospheric biosignatures by generating multiple lines of evidence to aid in assessing their biogenicity. Furthermore, surface biopigment features are the only way to detect more primitive forms of anoxygenic photosynthesis if the more metabolically complex oxygenic photosynthesis never evolved.  \textbf{\textit{Surface biosignatures instrument needs summary, key findings:} To detect the reflectance spectra of biological pigments on the surface of habitable exoplanets under atmospheric compositions that reflect different stages of Earth's history (Archean, Proterozoic, and Modern), an SNR 20-40 is needed in the visible to near-infrared wavelength range (~500-1100 nm). This is for 15\% total pigment coverage on abiotic surfaces under 50\% cloud cover. However, there may be some cases in which lower SNR is required; studies are ongoing.} \textbf{\textit{Coronagraph requirements} (1) The detection of surface biosignatures would be greatly enhanced by having as many parallel coronagraph channels as possible across the whole wavelength range with no or minimal gaps between channels. (2) Retrieval studies revealed that restricted wavelength ranges (e.g., 0.4 - 0.7 $\mu$m), such as may be used during initial survey strategies, are not sufficient to deconvolve the biopigment features from the abiotic background.}  
  \\
  \\
\end{abstract}

\section{Science Goal}

The fundamental question we aim to address in this HWO science case is: \textit{\textbf{How common are surface biosignatures of life in our solar neighborhood?}}

\subsection{Introduction}

\textit{Life's Requirements}. The search for life in the universe is rooted in our current understanding of the requirements and limits of life, and the observable traces or signatures that it leaves behind. This general set of requirements may be expected of all biological systems, with life on Earth being considered a specific expression within a particular environmental context. However, \textbf{life within and beyond our solar system may exhibit a range of solutions to these requirements, constrained by the laws of physics and principles of chemistry that presumably apply throughout the universe}. It is reasonable to design a life detection mission based on an Earthly use-case, but careful thought must be given to capabilities and trade-offs to enable a more agnostic search for life.

\begin{figure*}[!htb]
    \centering
    \includegraphics[width=0.4\textwidth]{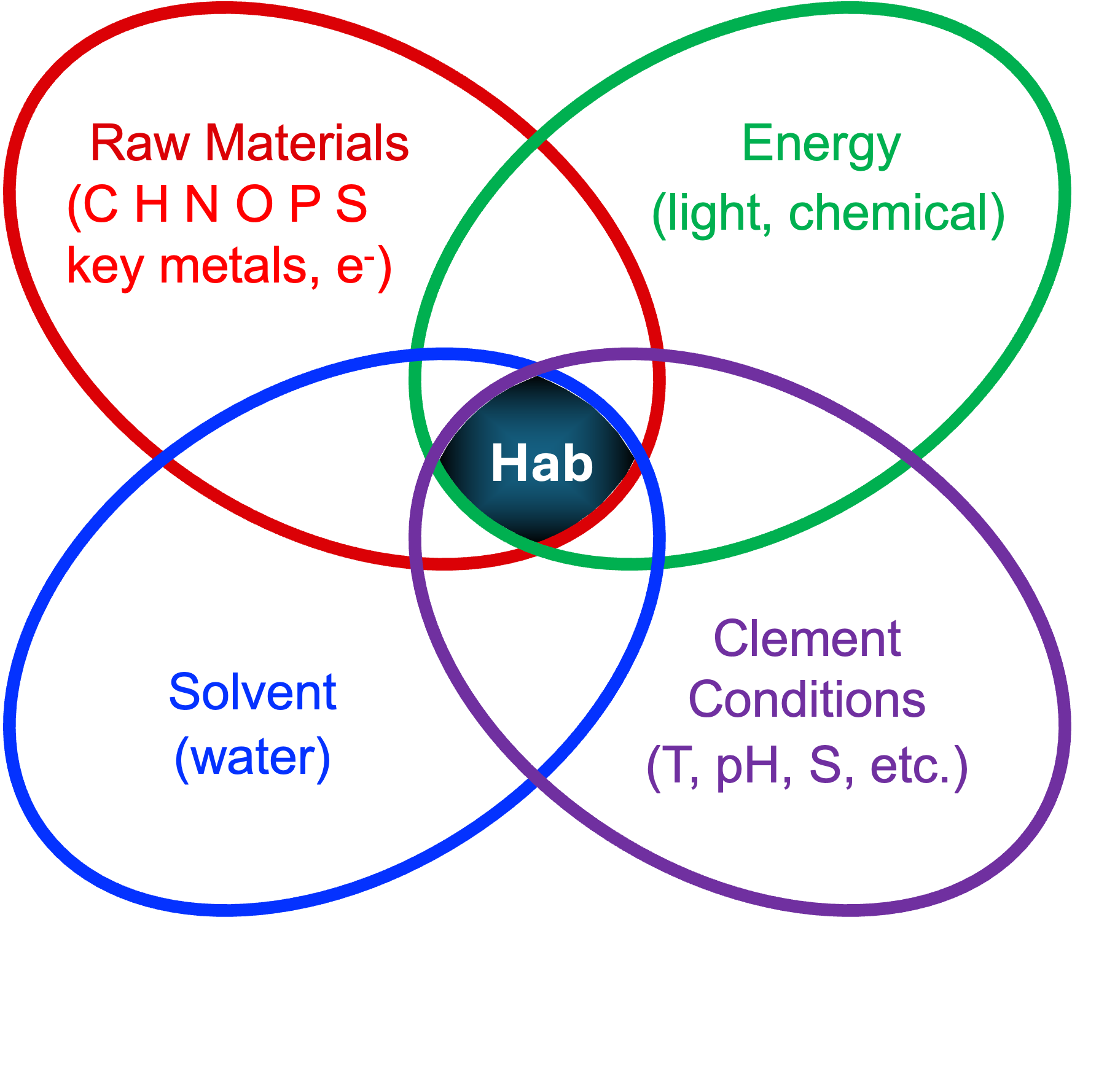}
    \caption{Requirements for life. An environment must provide these resources and conditions simultaneously in order to sustain habitable conditions (Hab) for life. Adapted from a graphic by Tori Hoehler.}
    \label{fig:fig1}
\end{figure*}

Generically, life requires a source of energy, a supply of elements with which to synthesize compounds (C, H, N, O, P, S and metals), a solvent to aid in their synthesis and molecular interactions, and "clement conditions" to allow for the full biochemical functionality of these molecules (Fig.~\ref{fig:fig1}). On Earth, energy requirements are met either by sunlight or by chemical reduction-oxidation (redox) reactions. Sunlight is a near-ubiquitous and plentiful energy source, whereas chemical reductants and oxidants are restricted to relatively small reservoirs on the planet. 

\textbf{Photosynthesis, the conversion of light energy to chemical energy to fix CO$_2$ into cellular materials, is an elegant solution to life’s requirement for energy.} Equation (1) generically represents photosynthesis where electrons from external donors are used in an energy conversion process to synthesize energy-rich molecules, which in turn are used in the "dark reactions" to fix CO$_2$ into cellular materials. In the equation, reductant = chemical species capable of donating electrons; oxidant = its oxidized equivalent; CH$_2$O = representation of organic compounds. In the case of oxygen-evolving (“oxygenic”) photosynthesis in plants, algae, and cyanobacteria, the reductant is H$_2$O, which is converted to O$_2$ via the removal of electrons and protons. Other reductants such as H$_2$, H$_2$S, and reduced iron or Fe(II) can be utilized in non-oxygen evolving (“anoxygenic”) photosynthesis. Anoxygenic photosynthesis is biochemically less complex and evolutionarily more ancient than oxygenic photosynthesis (Blankenship, 2010).

\begin{equation}
  \begin{aligned}
     Light + Reductant + CO_2 + 2H^+ \rightarrow  \\ 
    = Oxidized Product + CH_2O
  \end{aligned}
\end{equation}

\textit{Biological Pigments}. Details of the photosynthetic process can be found elsewhere, but briefly, it entails (1) light absorption, (2) charge separation, (3) electron transport, and (4) energy storage. Light absorption is achieved by pigments associated with protein complexes that are embedded in membranes within the cell. There are structural and chemical components of pigments that control the wavelengths of light that they absorb. \textbf{A key point to note is that the absorption maxima of pigments is flexible and can be tuned by the phototroph based on available light and environmental selection pressures.}

Photosynthetic pigments can be broadly grouped into two functional categories:  light-harvesting pigments and reaction center pigments.  Light harvesting pigments absorb light over a broad range of wavelengths and transfer the energy via resonance transfer \citep{Forster1967} to a reaction center that contains a chlorophyll or bacteriochlorophyll molecule in a special environment.  When the reaction center chlorophyll absorbs a quanta of light, an electron is excited and is donated or lost to an acceptor molecule.  This step is called charge separation and represents the first chemical change in the conversion of light energy to chemical energy. \textbf{It must also be noted that chlorophyll and bacteriochlorophyll pigments absorb across a range of wavelengths in the spectrum, and, for example, not just at the absorption maximum in the red we typically associate with Chlorophyll a (670-680 nm).} There are two pairs of absorption bands in the blue and red area of the spectrum, which are called the Q bands (Q$_x$ and Q$_y$) and B bands (or Soret bands) \textbf{(Table 1)}. There are also non-photosynthetic pigments within cells that function to, e.g., protect the cells from UV radiation or oxidative stress, detailed below. 
\vspace{12pt}

\begin{figure*}[ht!]
    \centering
    \includegraphics[width=0.9\textwidth]{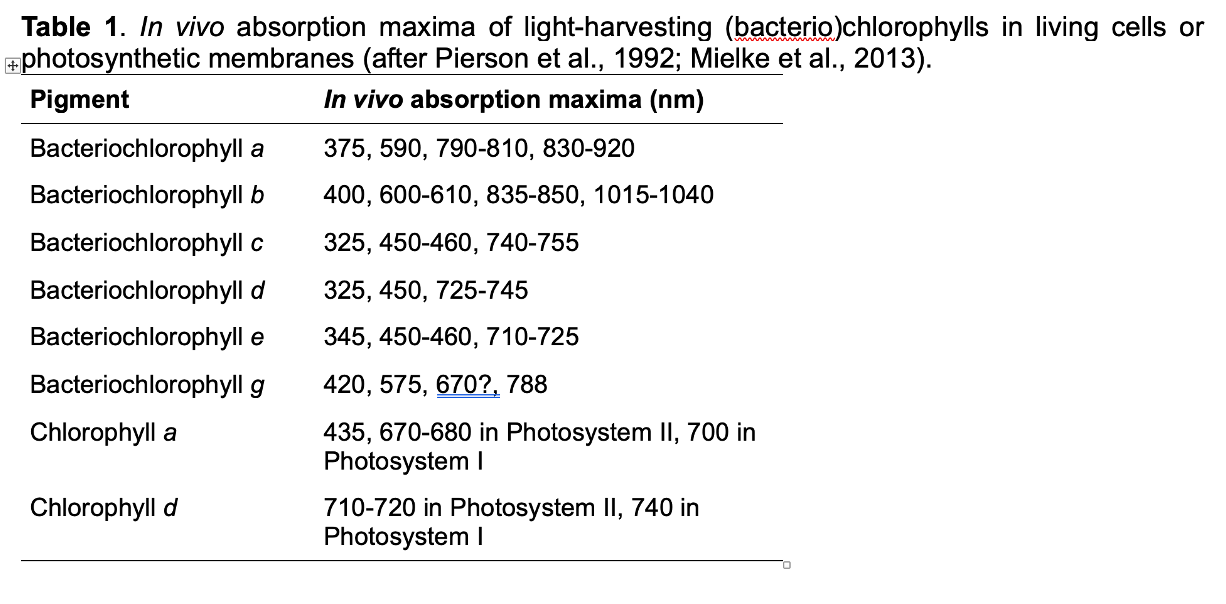}

\end{figure*}

\textit{Energy Flux and Exoplanets}. The flux of sunlight to the Earth provides such an abundant energy source that vegetation only absorbs 1-2\% of the total solar energy reaching the surface of the planet, and only 14\% of the incident flux in the 400-700 wavelength range \citep{Field1998}. Photosynthetic ecosystems have been estimated to dominate the gross primary productivity of the planet through the different stages in Earth's history \citep{crockford2023geologic}. Furthermore, the modern oxygenic photosynthetic biosphere is more productive by 3 or more orders of magnitude than was the chemosynthetic biosphere before the emergence of photosynthesis \citep[]{des2000did, canfield2006early}. 

\textbf{On rocky habitable exoplanets orbiting stars like our own, the available stellar photon flux suggests that photosynthetic biospheres, if present, would be more productive and detectable via their global-scale gaseous and surface pigment biosignatures than chemosynthetic biospheres.} On Earth, oxygenic photosynthesis is the most dominant type of metabolism in continental and marine habitats, and we can see remotely detectable evidence of it in Earth's spectrum via oxygen in the atmosphere and photosynthetic pigments covering the continents, the so-called "vegetation red edge" (VRE). Earthshine spectra from the Apache Point Observatory showed the VRE arising from the absorption and scattering of light by the photosynthetic pigment chlorophyll a \citep{seager2005vegetation}.

\subsection{Surface Biosignatures}
\begin{figure*}[ht!]
    \centering
    \includegraphics[width=0.9\textwidth]{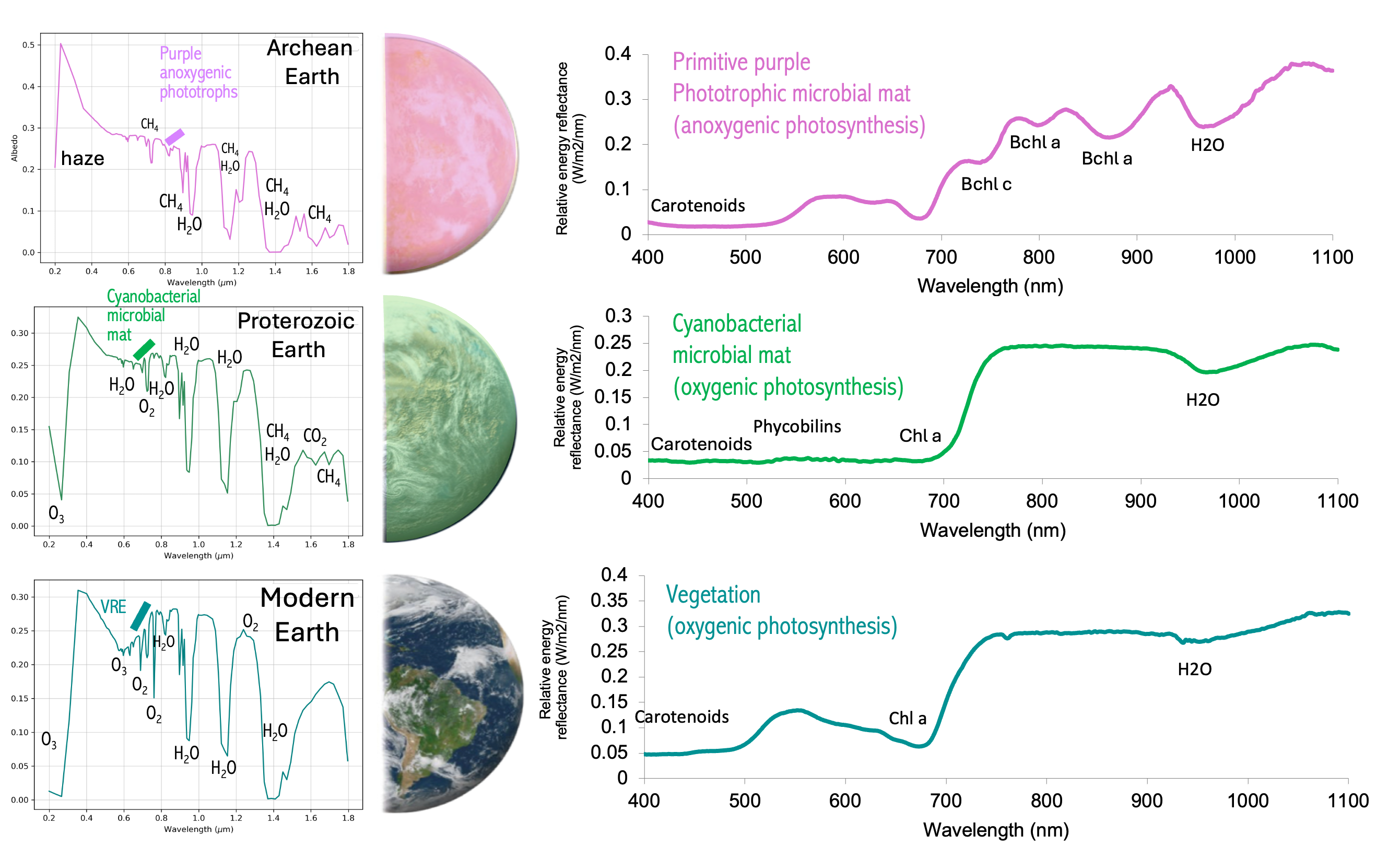}
    \caption{Left: Reflectance spectra of photosynthetic organisms on rock, mineral, and snow/ice abiotic surfaces with Archean, Proterozoic, and Modern atmospheric compositions (planetary spectra provided by Anna Grace Ulses, Univ. of Washington). The thick angled lines show the pigment "edge" features in the spectrum. Right: Top: purple anoxygenic phototrophs with bacteriochlorophyll pigments absorbance and reflectance features at ~805-880 nm at 15\% planet coverage under an Archean atmosphere with 50\% cloud cover. Middle: cyanobacterial microbial mat (oxygenic phototroph) with chlorophyll a absorption at ~660-680 nm at 15\% coverage under a Proterozoic atmosphere with 50\% cloud cover. Note that the 3-D structure of the mat is effective at scattering light to create a "red edge" feature similar to the vegetation red edge (VRE) shown in the bottom panel. Top two spectra provided by N. Parenteau, unpublished data. }
    \label{fig:fig2}
\end{figure*}

Surface biosignatures are planetary-scale spectral features resulting from absorption and/or scattering of radiation by organisms containing photosynthetic and non-photosynthetic pigments. The vegetation red edge is so called due to the sharp contrast in visible light absorbance by light harvesting pigments in plant leaves versus their high reflectance in the NIR (Fig.~\ref{fig:fig2}). This contrast occurs around 700 nm in the red/far-red. This reflectance feature is so distinct from Earth’s mineral spectra that it is used to detect vegetation cover with Earth-observing satellites \citep{tucker1985african}. In water bodies, chlorophyll a concentration can be obtained through measures of greenness \citep[]{o1998ocean, gohin2002five}.

\subsection{Rationale for Science Goal}
It's logical to speculate that photosynthesis could evolve on other planets, no matter whether it's primitive or more complex like oxygenic photosynthesis. Thus, it's reasonable to search for global scale spectral signatures of photosynthesis in the form of gaseous byproducts that can build up in the atmosphere (e.g., O$_2$ is a waste product from the use of H2O as an electron donor to ultimately fix CO$_2$ into CH$_2$O), or in the form of surface pigment biosignatures like the canonical vegetation red edge. \textbf{A key point to note is that the simpler forms of photosynthetic life present on the early and modern Earth do not generate volatile products when using electron donors such as H$_2$, H$_2$S, and Fe(II), and therefore their presence is only revealed by their photosynthetic pigments.}   

Oxygen as a biosignature has been addressed in the main Living Worlds Science Case Development Document. Here we propose a science case for the Habitable Worlds Observatory (HWO) to search in planetary reflectance spectra for absorption and reflectance features characteristic of surface pigments, which would provide evidence of photosynthesis or other non-photosynthetic processes that use biological pigments in other roles in the cell. This class of surface biosignatures, which is distinct from atmospheric biosignatures, can be used to corroborate volatile species such as oxygen and provide multiple lines of evidence to establish their biogenicity. These surface biosignatures can also possibly provide stand-alone evidence of life, especially for more primitive types of anoxygenic photosynthetic organisms. \textit{\textbf{Given the ubiquity of stellar energy in the universe and the possibility of the evolution of photosynthesis on other worlds, this leads us to ask the question "How common are surface biosignatures of life in our solar neighborhood?"}}

\section{Science Objective}

\textit{\textbf{Search for surface biosignatures on exoplanets and vet them in a false positive/false negative framework, or discover statistically meaningful limits for null results.}}

\vspace{10pt}
Measurements of reflected spectra from exoplanets may contain contributions from surface biosignatures if they are in the line of sight of the telescope. In real observations, we will have to compare these spectra to spectral properties of terrestrial (Earth) biological pigments (biopigments), as these will provide the only guidelines to determine their possible biogenicity. Such future observations will provide the first quantitative limits on the presence of photosynthetic life on exoplanetary surfaces. We may be able to infer surface maps of photosynthetic life on the planet if we have sufficiently resolved data over long-time baselines.

Because surface biosignatures are relatively underexplored compared to atmospheric biosignatures, the Surface Biosignatures Task Group aimed to perform a comprehensive assessment of published and unpublished data \textbf{(1) provide a spectral library of biological pigments and identify the absorption features, (2) provide an assessment of potential abiotic false positives from minerals, rocks, and snow/ice, and (3) perform numerical modeling of exoplanet reflectance spectra to constrain their detectability by HWO (addressed in the Physical Parameters and Description of Observation sections)}. For all three parts of these tasks, we build on the limited published results available in the literature to explore a wider parameter space than has been done in prior work. The spectral library compilation combs the literature and existing public/private data archives to catalog the large parameter space of different surface biosignatures which occur on Earth, and which may plausibly occur on exoplanets. These include pigments that may be adapted to a particular star (the so-called light harvesting pigments), as well as pigments that could have more universal functions. By establishing the spectral diversity of life on Earth, we hope that we can develop strategies for detecting surface biosignatures on alien worlds. 

\subsection{Earth Reference Cases for Surface Biosignatures}
\textit{Oxygenic photosynthesis - chlorophyll pigments}. Using Earth as our only example, detectability of chlorophyll a \textbf{(Table 1)} via the vegetation red edge has been investigated in spectra of Earthshine reflected from the Moon\citep[]{arnold2002test, woolf2002spectrum, rodriguez2004earthshine, montanes2005globally, montanes2006vegetation, seager2005vegetation, arnold2008earthshine}, and via spectropolarimetry \citep[]{sterzik2012biosignatures}, direct spacecraft observation of Earth \citep[]{sagan1993search, livengood2011properties}, with Earth radiance simulators \citep[]{tinetti2006detectability, fujii2010colors, kawahara2010global, robinson2011earth, fujii2012mapping}, and through different points in Earth’s history \citep[]{kaltenegger2007spectral, arnold2008earthshine}. In addition, oxygenic photosynthesis exhibiting potential pigment adaptations to alternative parent star spectra have been proposed \citep[]{wolstencroft2002photosynthesis, kiang2007spectral}, and a hypothetical “NIR edge” for vegetation adapted to M stars was found possibly to be more easily detectable through an Earth-like atmosphere than the vegetation red edge \citep{tinetti2006detectability}.

\vspace{10pt}
\textit{Anoxygenic phototrophs - bacteriochlorophylls pigments}. Oxygenic phototrophs are the primary drivers of carbon fixation on modern Earth; however, anoxygenic phototrophic bacteria, which utilize bacteriochlorophyll pigments (BChls), dominated primary productivity on the early Earth before the evolution of oxygenic photosynthesis. These bacteria are classified into six major phyla: Proteobacteria (purple photosynthetic bacteria), Chlorobi (green sulfur bacteria), Chloroflexi (filamentous anoxygenic phototrophs [FAPS]), Firmicutes (heliobacteria), Acidobacteria, and Gemmatimonadetes \citep[]{bryant2007candidatus, zeng2014functional}. Each lineage is distinguished by unique photosynthetic apparatuses, with variations in light-harvesting complex architecture and carotenoid composition, but all rely on BChls for photon capture.

Some of these bacteria, termed aerobic anoxygenic phototrophs (AAP), are not exclusively photosynthetic. AAP bacteria produce BChl a under fully oxic conditions and can constitute 1–7\% of total prokaryotes in oligotrophic ocean regions. Their abundance increases to 2–15\% in coastal and estuarine environments, highlighting the adaptability of BChl-producing organisms \citep{koblivzek2015ecology}.

Anoxygenic phototrophs are believed to have evolved in surface environments during the Archean eon when Earth’s atmosphere was predominantly anoxic. These phototrophs were used in Archean Earth models to assess the detectability of surface biosignatures on the Archean "purple Earth" e.g.,\citep{ sanroma2013characterizing}. In modern environments these phototrophs are sensitive to oxygen and typically live beneath surface oxygenic chemotrophs in chemoclines where a flux of reductants intersects with low oxygen levels and available light. Notable exceptions include surface purple microbial mats, which grow beneath a diffusive oxygen barrier, such as those observed in the Sippewissett Marsh in Massachusetts \citep{kosmopoulos2023horizontal}, amplifying the target environments in which these bacteria can thrive and potentially dominate. 

\begin{figure*}[ht!]
    \centering
    \includegraphics[width=0.9\textwidth]{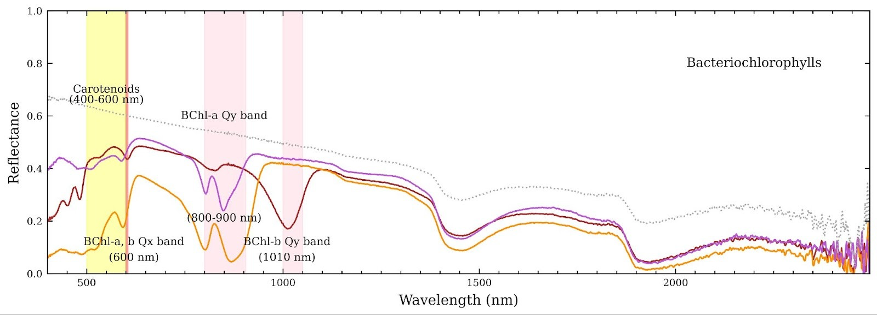}
    \caption{Reflectance spectra of bacteriochlorophylls a and b (red), as well as carotenoids (yellow) from purple non-sulfur bacteria. Note absorption near-infrared features from Bacteriochlorophylls span between 800-1010 nm. Data from Coelho et al., 2024.}
    \label{fig:fig3}
\end{figure*}

The BChls utilized by anoxygenic phototrophs enable light capture across a broad wavelength range, from 400 nm to 1000 nm (Fig.~\ref{fig:fig3}). The longest wavelength known to drive photosynthesis—1018 nm—has been observed in the purple bacterium \textit{Blastochloris viridis}, which may make it well suited for life on an M-dwarf hosted planet \citep[]{wolstencroft2002photosynthesis, kiang2007spectral, coelho2024purple}. \cite{metz2024detectability} found that if a \textit{Blastochloris viridis}-like organism were present on Proxima Centauri-b in moderately large quantities, its impact on the planetary reflectance spectrum would be detectable by LUVOIR-A/B, and maybe on other exoplanets within 15 pc. The physiological and biochemical properties of anoxygenic phototrophs vary widely among species, giving rise to distinct surface biosignatures \citep{coelho2024purple}.

\vspace{10pt}
\textit{Carotenoids}. Carotenoids pigments are widely distributed in nature, being common in bacteria, archaea, plants, and algae. They represent a broad spectrum of colors (e.g., red, yellow, purple, orange, pink) and comprise more than 1,117 distinct chemical profiles \citep{yabuzaki2017carotenoids}. Their presence across all major life forms is a testament to their versatility, not only in light harvesting for photosynthesis \citep[]{vogl2011elucidation, vogl2012biosynthesis} but also in enabling organisms to survive in extreme environments. For instance, carotenoids provide photoprotection through mechanisms such as the xanthophyll cycle, protection against dryness by influencing membrane fluidity and thermal photoprotection by stabilizing other pigments \citep[]{edge1997carotenoids, seel2020carotenoids}.

Carotenoids are terpenoid compounds typically consisting of 40 carbon atoms, some of which may include aromatic rings. Aromatic carotenoids, in particular, are important markers of oxygen evolution on Earth and serve as signatures of euxinic environments (low oxygen, high sulfur), where they are commonly found in phototrophic sulfur bacteria \citep{ma2022aromatic}.

\begin{figure*}[ht!]
    \centering
    \includegraphics[width=0.5\textwidth]{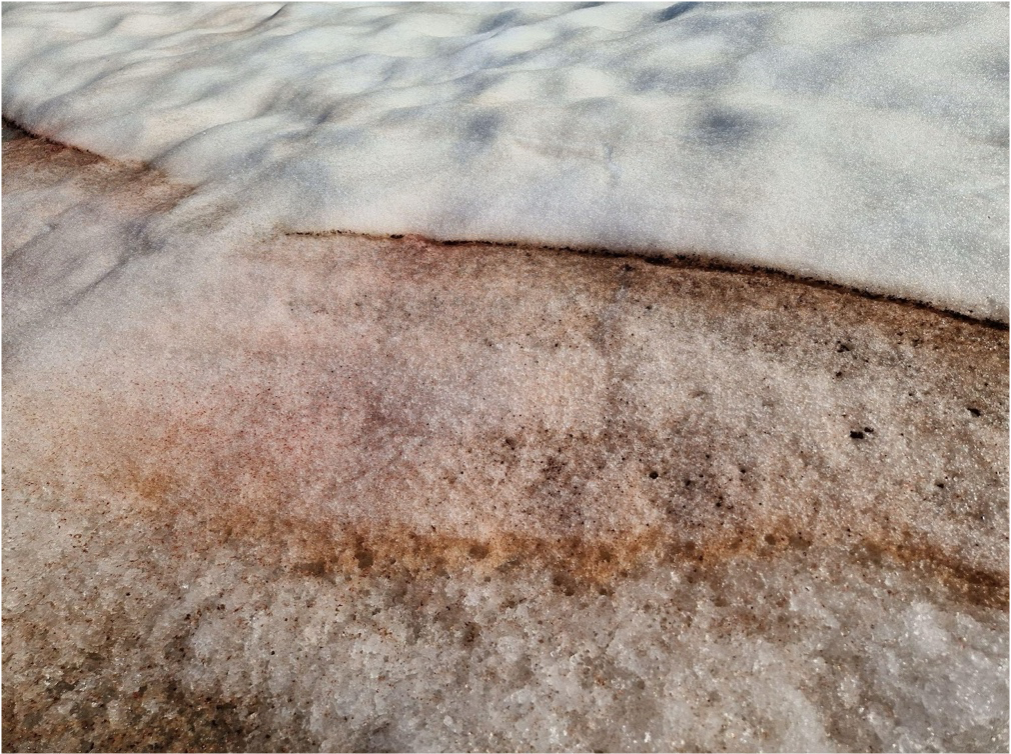}
    \caption{Glacier algae carotenoids covering the surface of Glacier Austre Brøggerbreen (Svalbard), image by Ligia F. Coelho (Cornell University).}
    \label{fig:fig4}
\end{figure*}

These pigments are characterized by specific and identifiable spectral fingerprints and exhibit remarkable resistance to irradiation and desiccation, making them favored molecules in polar and desert ecosystems \citep{vitek2017discovery}. A prime example is the diverse array of carotenoids produced by ice and snow biota, where they play crucial roles in both photoprotection and light harvesting \citep{coelho2022color}. Some of these pigments, as shown in (Fig.~\ref{fig:fig4}), can even be detected through remote airborne imaging \citep{painter2001detection}.

\vspace{10pt}
\textit{Bacteriorhodopsin (retinal pigments)}. Unlike chlorophyll, bacteriorhodopsin functions as a light-driven proton pump, generating energy by absorbing green light (~570 nm) while displaying spectral complementarity with chlorophylls and bacteriochlorophylls. However, rhodopsin-based phototrophy is not observed to be linked to carbon fixation (i.e., it allows energy production but not biomass production). \cite{dassarma2021early} proposed that this spectral complementarity might hint at the evolutionary timeline of rhodopsins relative to chlorophyll pigments and present an alternative surface biosignature for exoplanet exploration. \cite{sephus2022earliest} expanded on this idea, suggesting that microbial rhodopsins, which likely originated as green-light-absorbing proton pumps, evolved in response to Earth's early photic conditions. This implies that inhabited exoplanets could exhibit diverse spectral patterns that may include analogs to rhodopsins. 

\vspace{10pt}
\textit{Emissive surface biosignatures: Phosphorescence, fluorescence, bioluminescence}. Both vegetation and microorganisms can emit light via fluorescence and bioluminescence \citep[]{papageorgiou2007fast, haddock2010bioluminescence}. Chlorophyll autofluorescence occurs when pigments absorb UV light and re-emit it at lower energy levels, primarily in the red spectral range (640-800 nm) with peaks at 685 and 740 nm. Although these signals are faint, they are detectable through high-resolution spectroscopy from satellites \citep[]{joiner2011first, sun2017oco}. Some researchers suggest that biological autofluorescence could serve as a transient surface biosignature, especially in response to stellar flares \citep{o2018biofluorescent, o2018vegetation}. \cite{komatsu2023photosynthetic} analyzed the potential to detect autofluorescence biosignatures on exoplanets, concluding that bacteriochlorophyll fluorescence (1000-1100 nm) could be observed on planets around ultracool dwarfs using telescopes similar to HWO, as these wavelengths align with absorption features in the stellar spectrum. Possible false positives include minerals like fluorite and calcite, detectable from space, though their fluorescence profiles differ from biological signals \citep{kohler2021mineral}.

In contrast, bioluminescence is the active emission of light, produced by organisms via the oxidation of luciferins. It is found across diverse life forms, including bacteria, protists, and animals \citep{haddock2010bioluminescence}. Some marine microbes emit bioluminescence detectable by Earth satellites, forming patches over 10,000 km² \citep{miller2005detection}, leading to its suggestion as a potential biosignature for exoplanets \citep{seager2012astrophysical}. However, the low overall brightness of these patches means their detectability on exoplanets remains unquantified

\vspace{10pt}
\textit{Land vs. marine biota signals}. The vegetation red edge (VRE) refers to the contrast between the absorbance of chlorophylls which occurs in the red (at 670 nm for chlorophyll a) versus the high reflectance of plant leaves in the near-infrared (NIR).  The VRE is a distinct signature from mineral spectral and has been widely used to monitor Earth’s land vegetation (Tucker et al. 1979). Higher chlorophyll absorbance occurs for thicker leaves, so plants like cacti would generate significantly larger leaf-level VRE than shrubs or grasses \citep{slaton2001estimating, arnold2008earthshine}. Additionally, ‘land-ocean planets’ with higher land fractions than oceans and with continental interior climates would have fewer clouds and higher surface albedos on global average, making VRE potentially easier to detect on these worlds, depending on potential density of the vegetation \citep{arnold2008earthshine}. Consequently, land-ocean desert worlds with extensive land biospheres supported by large, vascular plants may represent ideal planetary candidates for using VRE as a life detection tool.  In the marine biosphere, chlorophylls can be observed remotely through quantification of ocean color. Q bands (VRE) are expected to be muted but not through red-edge contrasts in reflected light signals due to the low albedo of water in the NIR at these wavelengths \citep{o1998ocean, seager2005vegetation}. Chlorophylls have other high intensity peaks in the blue region of the spectrum called Soret bands that are more favorable to be searched for in marine environments \citep{chowdhary2019modeling}.  In seasonally dry environments, dry biopigments can produce significantly more intense signals than wet \citep{coelho2022color}.

\vspace{10pt}
\textit{Temporal variations in surface biosignatures (‘seasonality’)}.
The strength of VRE signals fluctuate seasonally across each hemisphere on Earth \citep{miller1991seasonal}. We therefore expect that a similar seasonality in the VRE signal could arise on exoplanets with vegetation. If VRE seasonality is detectable on exo-Earths, it could be exploited in several ways. For instance, certain abiotic spectral signatures that could mimic VRE in remote observations exist (e.g., cinnabar and sulfur minerals; \citep{seager2005vegetation}, but observing seasonal variations in the intensity of the reflectance peak may help rule out these potential false positives. It is also possible that ‘exo-plants’ use pigments other than chlorophyll to absorb stellar light, which may change the shape of the reflectance peak signal and/or shift it to a different wavelength \citep{schwieterman2018exoplanet}. In that case, detecting seasonal variations in a similar ‘edge-like’ spectral feature may provide clues that the signal observed is driven by life. It should be noted though that the seasons can also change the disk-integrated colors through the cloud patterns and snow coverage, and differentiating biological seasonality from these non-biological processes will also be needed.   

\vspace{10pt}
\textit{Green Ocean as an indicator of Early Surface Oxidation Driven by Phototroph Metabolism}. Earth has coevolved with life since its emergence on the surface approximately 4.0 Ga. A surface biosignature can be interpreted as global changes in the surface environment induced by biological activity. This allows signals related to life to be extracted from planetary reflected light, which is primarily shaped by geological features. A strong indicator of Earth's coevolution with life is the oxidation of its surface environment. Earth has gradually undergone oxidation over approximately 3.0 Gyr, starting with the emergence of oxygenic photosynthetic organisms such as cyanobacteria \citep{holland2006oxygenation}. Geological records, including banded iron formations (BIFs; \cite{konhauser2017iron}, Stromatolite reef in the early Archean era \citep{allwood2006stromatolite}, and the evidence of shallow-water oxidation before the Great Oxidation Event (GOE) ~2.4 Ga \citep{planavsky2014evidence}, suggest that surface oxidation began in aquatic environments due to the metabolic activity of photoferrotrophs and cyanobacteria.

\cite{matsuo2025archaean} propose a coevolutionary relationship between oxygenic phototrophs and their underwater environments during the Archean and Proterozoic eras. The underwater light environment, limited to green wavelengths by the oxidation of reduced iron driven by photoferrotroph and cyanobacterial metabolism, likely influenced the light-harvesting antennae of cyanobacteria. This change in underwater transmission spectra, caused by the presence of iron oxides, suggests that the color of the ocean shifted from blue to green following the emergence of phototrophs.

\begin{figure*}[ht!]
    \centering
    \includegraphics[width=0.9\textwidth]{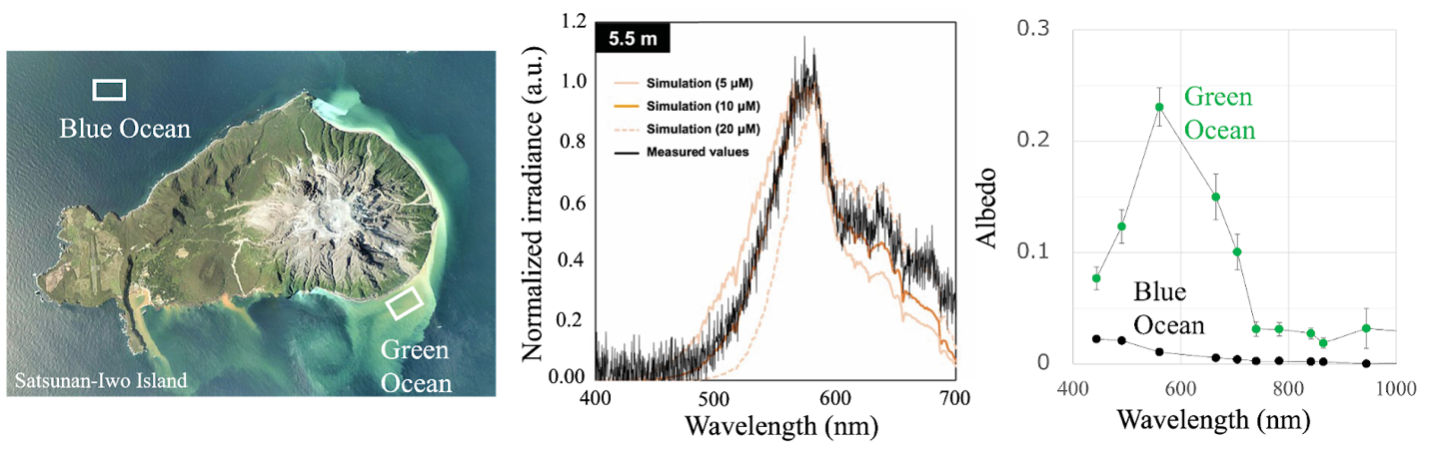}
    \caption{Aerial photograph of Iwo island (left), transmitted light spectrum measured at the depth of 5 m (center), and comparison of albedos between a typical blue ocean and a green ocean. Adapted from Matsuo et al., 2025.).}
    \label{fig:fig5}
\end{figure*}

Modern analogs to the Archean and Proterozoic aquatic environments exist in hydrothermal vent systems, such as those around the Satsunan-Iwo islands in Kyushu, Japan \citep{kiyokawa2021hydrothermal}. Through remote sensing and field research, \cite{matsuo2025archaean} confirmed a direct relationship between underwater transmission spectra and the observed green coloration of the ocean in these environments (Fig.~\ref{fig:fig5}). Notably, the presence of iron hydroxides not only restricts the underwater light spectrum to green but also alters the ocean’s color to green. Interestingly, they observed a significantly higher albedo in the 500–600 nm wavelength range compared to that of a typical blue ocean (Fig.~\ref{fig:fig5}).

These findings suggest that the green ocean may represent the earliest oxidation event of Earth's surface environment, driven by phototrophic metabolism before the GOE. The intermittent occurrence of green oceans persisted throughout the Archean and Proterozoic eras, up to the Neoproterozoic Oxidation Event (NOE). Although some iron oxides may have formed through UV-induced oxidation prior to the GOE, this mechanism is not considered the primary driver for BIF formation \citep{konhauser2007decoupling}. Instead, the metabolic activity of photoferrotrophs and cyanobacteria is likely the dominant driver of green ocean formation.

\vspace{10pt}

\begin{figure*}[ht!]
    \centering
    \includegraphics[width=0.6\textwidth]{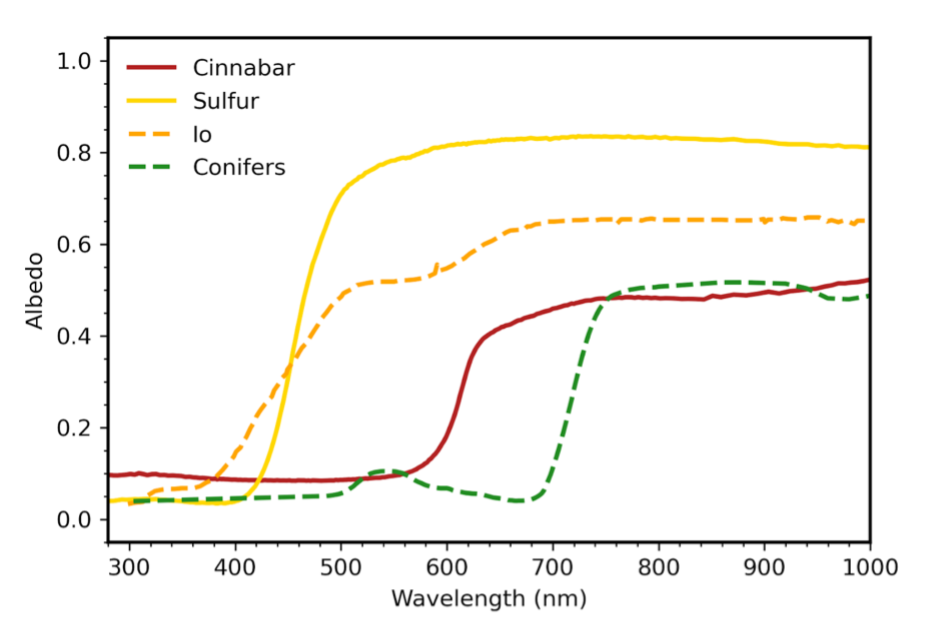}
    \caption{Comparison of VRE and false positive mineral “edge” features. Shown are the reflectance spectra of elemental sulfur and cinnabar (HgS) sourced from the USGS spectral library (Clark et al. 2007), the reflectance spectrum of the jovian moon Io from Karkoschka (1994), and a spectrum of a conifer forest from the ASTER spectral library (Baldridge et al. 2009). Adapted from Schwieterman (2018). }
    \label{fig:fig6}
\end{figure*}

\textit{Abiotic False Positives for Reflectance Biosignatures}. Pigments and other manifestations of life are not the only possible surface properties of a planet that may produce “edge”-like spectral features. Electronic transitions in mineral semiconductors can also generate sharp rises in spectral albedos analogous to the vegetation red edge \citep{seager2005vegetation}, albeit at alternative wavelengths. These minerals will create edge features at wavelengths that can be estimated based on their intrinsic band gap energies, which are measured from their absorption or reflectance spectra. (Fig.~\ref{fig:fig6}) shows the spectral albedos of elemental sulfur, cinnabar (HgS, mercury (II) sulfide), and Jupiter’s moon Io (which contains sulfur and other minerals), contrasted with a conifer forest for comparison. We note that elemental sulfur and cinnabar themselves are unlikely to persist in large quantities on the surface of a habitable planet with a robust hydrological cycle. Therefore, as always, context will be essential in correctly interpreting surface biosignatures and ruling out potential false positives. To date, no VRE-specific (i.e., ~700 nm edge) mineral mimic has been proposed.

\section{Physical Parameters}

Astro2020 suggests that HWO examine ~25 terrestrial planets in their stars’ habitable zones for signs of habitability and life. This sample size would guarantee seeing at least one planet with biosignatures at 95\% confidence, if the frequency of inhabited planets with observable global biospheres is ~10\% of all candidates (Fig.~\ref{fig:fig6}). The main Living Worlds SCDD additionally tied the sample size to the "Earth through Time" approach and supposed that if 10\% of rocky habitable zone planets represent an Earth twin at a random point in Earth’s evolution, then 33 planets must be observed to detect biosignatures at 95\% confidence because 11\% of the “Earths” in the sample would be Hadean-like and uninhabited.  

\begin{figure*}[ht!]
    \centering
    \includegraphics[width=0.4\textwidth]{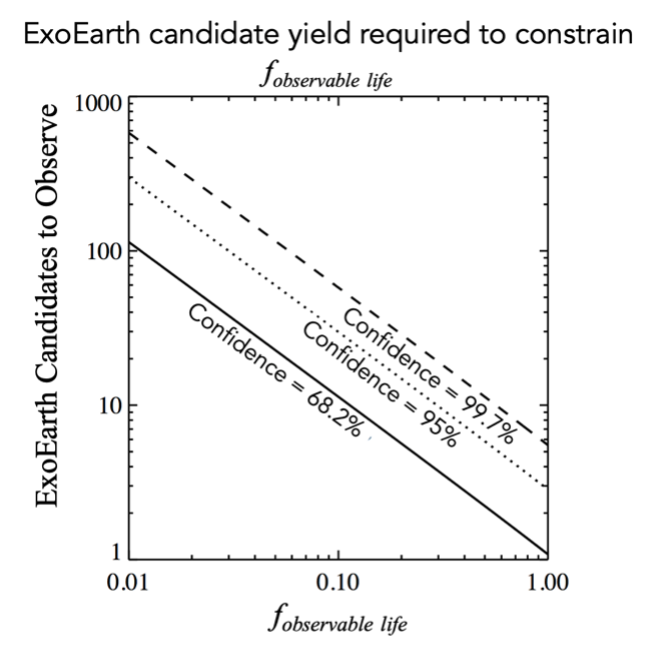}
    \caption{The number of candidate planets that must be observed to constrain various fractions of planets with observable life/biosignatures for given confidence levels. Credit: Chris Stark }
    \label{fig:fig7}
\end{figure*}

More generally, the number of candidate exoplanets ($N_{ec}$) required to constrain the fraction of planets with a given characteristic x ($\eta_{x}$) at a given confidence level (c) can be written as:
\begin{equation}
    N_{ec}=log(1-c)/log(1-\eta_{x}). 
\end{equation}

We can visualize what this means in Fig.~\ref{fig:fig7}. It is also useful to think about what this means in the case of a null detection. If we examine 25 candidate spectra and do not see signs of life, then we can say that the frequency of habitable planets with observable signs of life is $<$ 10$\%$ of candidate planets in the nearby universe at 95$\%$ confidence, placing the first ever upper limit on the frequency of observable biospheres in the cosmos. 

\begin{figure*}[ht!]
    \centering
    \includegraphics[width=0.9\textwidth]{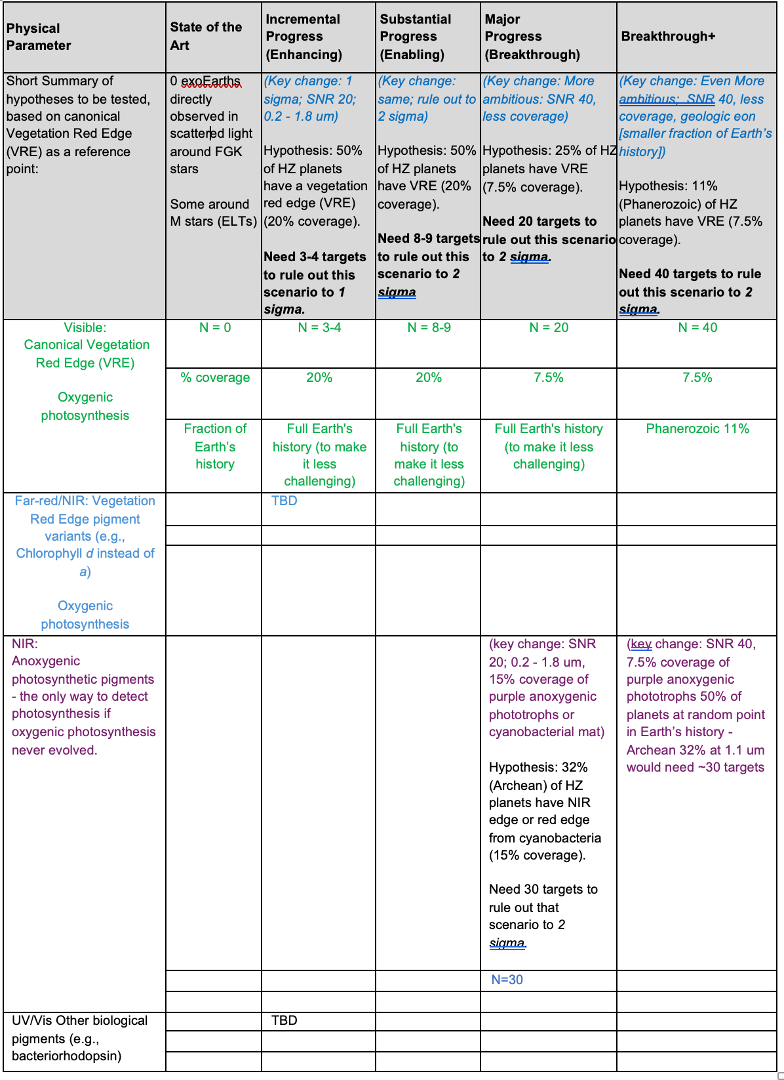}
    \caption{\textbf{Table 2.} The number of targets for various surface biosignatures hypotheses to be tested to various science return levels. These are based off the canonical vegetation red edge (VRE) and then expands to include the more primitive anoxygenic phototrophs 'NIR edges' for the Major Progress and Breakthrough+ science return levels. This would allow detection of photosynthesis even if oxygenic photosynthesis had never evolved. These science return levels can be reassessed.  }
    \label{fig:fig8}
\end{figure*}

\textbf{For surface biosignatures, we computed the number of targets needed to test various hypotheses regarding the detectability of the VRE and anoxygenic phototrophs at a variety of surface coverages and confidence levels for various science return levels \textbf{(Table 2; Fig.~\ref{fig:fig8})}. The 'Major Progress' level was designed to characterize ~25 exoplanets recommended by the Astro2020 Decadal Survey \citep{lancaster2018decadal}. We used the canonical VRE as our key use case given the extensive studies in the literature, and the general familiarity of the astrophysics and planetary communities with this surface feature. We included the more primitive anoxygenic phototrophs 'NIR edges' in the 'Major Progress' and 'Breakthrough+' science return levels. Seeking signs of anoxygenic photosynthesis via surface pigmentation would allow for detectoin of global photosynthetic biospheres even if oxygenic photosynthesis (and the corresponding atmospheric O$_2$ biosignature) never evolved.}

\section{Description of Observations}

\textit{Spectral resolution}.  As with any spectral measurements, there is a tradeoff between high resolving power and precision of the radiance measurement at each resolution element. In other words, lower resolution spectra have less noise and vice versa. Thus, we are faced with the need to resolve the surface biosignature spectra feature while maintaining a sufficient SNR to detect the feature above the noise. Investigations into this tradeoff are ongoing. Preliminary results indicate that a resolving power less than 130 would be optimal for detecting the VRE contrast, but further study is needed to see what lower bound still allows for differentiation between mineral mimics. Speaking in general from first principles, a resolution of 10 nm would be sufficient for most pigment edge-like features and not a challenge for design of HWO.

\begin{figure*}[ht!]
    \centering
    \includegraphics[width=0.9\textwidth]{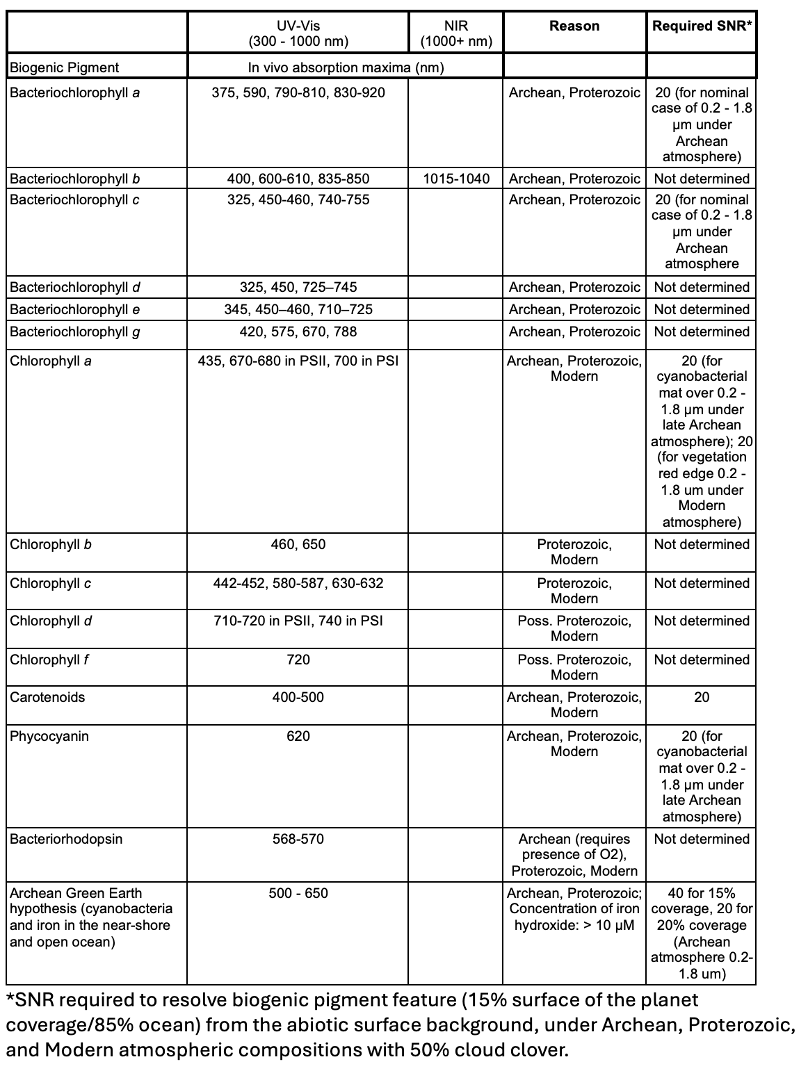}
    \caption{\textbf{Table 3.}Desired spectral features for the search for surface biosignatures. Recreated with permission from Edward Schwieterman. The information was sourced in part from: Schwieterman, E. W. and Leung, M. (2024) ‘An Overview of Exoplanet Biosignatures’, Reviews in Mineralogy and Geochemistry, 90(1), pp. 465–514. doi: 10.2138/rmg.2024.90.13.). The required SNR to detect the different biopigments over Earth's history is reported; results from retrival models shown in Figure 11.}
    \label{fig:fig9}
\end{figure*}

\textit{Signal-to-noise ratio (SNR)}.  The required SNR depends on both the intrinsic contrast in the \textit{in situ} reflectance signal and its spatial cover over the planet.  \textbf{We assessed the SNR required to detect surface photosynthetic pigments using the "Earth through time" framework (Table 3; Fig.~\ref{fig:fig9})}. We used the same atmospheric compositions for the Archean, Proterozoic, and Phanerozoic as were used in the retrieval models in the main Living Worlds SCDD. We self-consistently chose different biopigments at various surface coverage levels to reflect the photosynthetic populations present on the surface of the planet through its geologic eons, as indicated by evidence in the rock record. We tied our analyses to the canonical modern VRE to allow for easier comparison across the detectability scenarios to assess the possible scientific yields if such observations were performed by HWO. We tested some of the parameter space which may affect detectability, but ultimately this is a large parameter space which can be investigated more fully in future work. We also included assessment of cyanobacterial microbial mats that were present in the late Archean and Proterozoic, the Archean Green Earth hypothesis \citep{matsuo2025archaean}, as well as purple anoxygenic phototrophs.

\textit{Methods}. The spectral modeling part of this task involved generating synthetic planetary reflection spectra with radiative transfer models to observe the impact of surface biosignatures on planetary reflectance spectra. With these synthetic spectra, we conducted experiments to constrain the detectability of the surface features by adding simulated instrument noise to the spectra, to see if the absorption feature from the surface biosignature can be discerned in the spectrum. Our detection algorithm finds a best fit for the planetary properties, including atmospheric and surface composition, to reconstruct the synthetic spectra. If our algorithm found the best fit for the surface to include a biopigment in an abundance that is close to the true value that we used when generating the synthetic spectrum, we considered that to be a detectable scenario. In other words, this scenario would lead to an observable impact on the planetary reflectance spectrum, that HWO could plausibly detect. 

\begin{figure*}[ht!]
    \centering
    \includegraphics[width=0.9\textwidth]{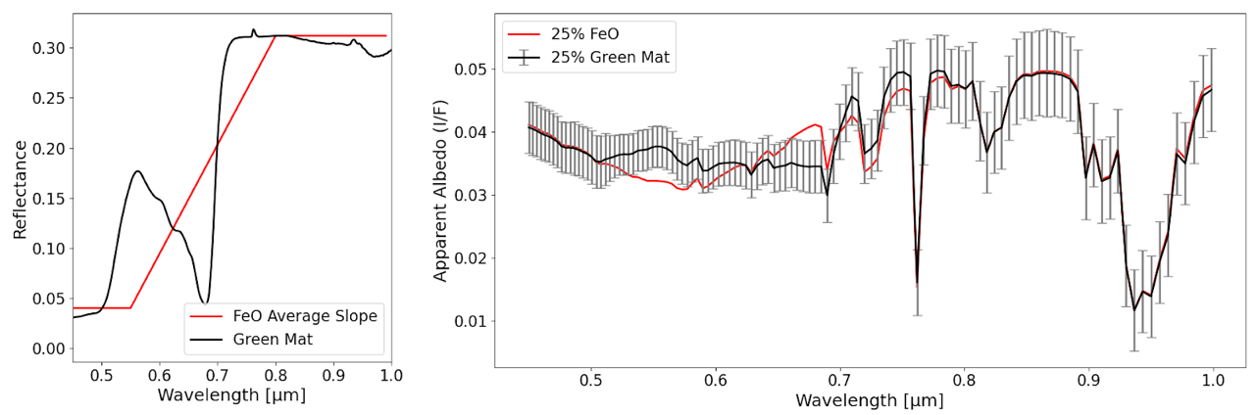}
    \caption{Left. Black plot: reflectance spectrum of a microbial mat composed of cyanobacteria, which are chlorophyll a-containing oxygenic phototrophs (aka "green mat"). The steep increase in reflectivity at 0.7 um is known as the vegetation red edge feature and is also displayed by cyanobacterial mats. There is also an increase in reflectance from 0.5 - 0.7 um due to absorption by carotenoids in the blue and absorption by chlorophyll a in the red; the rest of the light is scattered. Spectrum adapted from Borges et al., 2024. Red plot: reflectance spectrum of an iron oxide, which is an inorganic mimic for the vegetation red edge, but with a gentler slope (spectrum from Borges et al., 2024). Right. Two synthetic reflectance spectra of exoEarths at 10pc, one inhabited (black plot) and one uninhabited (red plot), observed with an HWO-like telescope. The black spectrum is a planet with a surface coverage of 25\% green mat with simulated instrument noise shown as gray error bars. The red spectrum is an identical planet except the surface is composed of 25\% iron oxide. With this integration time, the noise would be low enough so that the increased reflectance from 0.5 - 0.7 um could be detected. The red edge feature could be potentially differentiated from the gentler upslope of FeO from 0.6 - 0.7 m. This figure is only included as an illustration of our methods, and not meant to be a quantitative result.  }
    \label{fig:fig10}
\end{figure*}

An example of the spectral modeling that was performed is shown in Fig.~\ref{fig:fig10}, where the reflectance spectrum of an inhabited planet is contrasted with an uninhabited planet to perform an assessment of abiotic false positives. The reflectivity of the mineral iron oxide mimics the vegetation red edge over the same wavelength range, although the slopes differ. Any potential detection of a surface biopigment must include a thorough vetting of all known mineral and rock spectra contained in e.g., the United States Geological Survey (USGS) Spectral Library. 

\begin{figure*}[ht!]
    \centering
    \includegraphics[width=0.9\textwidth]{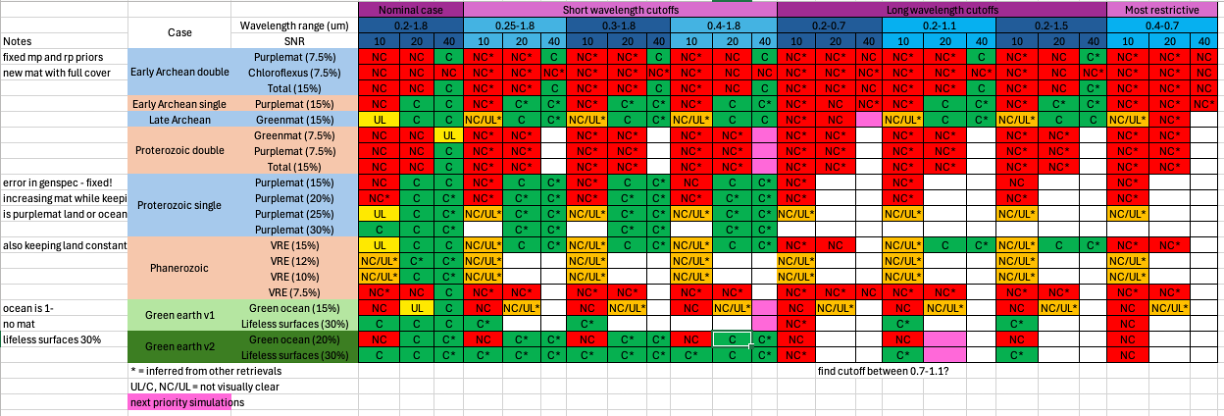}
    \caption{Spectral retrievals of the NIR edges of anoxygenic phototrophs, red edge of cyanobacterial microbial mats, "Archean Green Earth," and vegetation red edge from different abiotic surface types at atmospheric compositions over Earth's history (Credit: Anna Grace Ulses, University of Washington).}
    \label{fig:fig11}
\end{figure*}

\textit{Results}. \textbf{A detailed summary of the retrievals performed using the of 'Earth through time' framework is shown in Fig.~\ref{fig:fig11}}. For the early Archean eon over the full HWO wavelength range ($\sim$0.2 - $1.8 \text{\textmu}m$), the purple anoxygenic phototrophic mat (15\% coverage) required an SNR of 20 to deconvolve it from the abiotic background. Studies examining whether the spectral range could be lessened at either the short or long wavelength cut-offs, or both, revealed the detectability decreased and an SNR of 40 was required. The cyanobacterial microbial mat (oxygenic phototroph; 15\% coverage) revealed a similar result. Interestingly, the Archean Green Earth hypothesis (cyanobacteria living in the upper water column of an iron-rich ocean) \citep{matsuo2025archaean} revealed that the surface features were detectable at 15\% coverage with an SNR of 40. However, the 15\% surface coverage is a modest estimate given the lateral extent of the ocean, and increasing coverage to 20\% decreased the SNR to 20. It's reasonable to expect even more substantial coverage levels across near-shore areas and parts of the open ocean, which would decrease the SNR further to 10 or even 5. \textbf{These cyanobacterial oxygenic phototrophs living in iron-rich surface waters may be the most detectable surface biosignature on an Archean Earth} \citep{matsuo2025archaean}.

For the Proterozoic eon, the purple anoxygenic phototrophs revealed trends similar to the Archean results at 15\% coverage. Optimistically, when the coverage level was increased to 30\%, the required SNR to detect the pigments through the atmosphere, clouds, and on the abiotic land surfaces decreased to 10. Experiments with two different types of microbial mats representing realistic microbial communities yielded more sobering results. Two different mat types (purple bacteria and cyanobacteria) had multiple red and NIR edges, which made them difficult to detect. 

For the Phanerozoic (Modern) eon, the VRE required the full wavelength range (0.2 - $1.8 \text{\textmu}m$) to be detectable at 15\% coverage levels at an SNR of 20. Experiments moving the short wavelenth cut-off red-ward did not substantially change detectability given that there are no strong VRE features shortward of 600 nm (0.25 - $1.8 \text{\textmu}m$, 0.3 - $1.8 \text{\textmu}m$, and 0.4 - $1.8 \text{\textmu}m$). Not surprisingly, moving the long wavelength cut-off blue-ward demonstrated that the VRE pigment features could not be constrained from the abiotic surface types (0.2 - $1.5 \text{\textmu}m$, 0.2 - $1.1 \text{\textmu}m$, and 0.2 - $0.7 \text{\textmu}m$. And using a very restricted wavelength range (0.4 - $0.7 \text{\textmu}m$), such as may be used in initial survey strategies, revealed that the VRE would not be detectable.

\section{Concluding Remarks}
\textbf{\textit{Surface biosignatures instrument needs summary, key findings:} To detect the reflectance spectra of biological pigments on the surface of habitable exoplanets under atmospheric compositions that reflect different stages of Earth's history (Archean, Proterozoic, and Modern), an SNR 20-40 is needed in the visible to near-infrared wavelength range (~500-1100 nm). This is for 15\% total pigment coverage on abiotic surfaces under 50\% cloud cover. However, there may be some cases in which lower SNR is required; studies are ongoing.} 

\textbf{\textit{Coronagraph requirements} (1) The detection of surface biosignatures would be greatly enhanced by having as many parallel coronagraph channels as possible across the whole wavelength range with no or minimal gaps between channels. (2) Retrieval studies revealed that restricted wavelength ranges (e.g., 0.4 - $0.7 \text{\textmu}m$), such as may be used during initial survey strategies, are not sufficient to deconvolve the biopigment features from the abiotic background.} 

\bibliography{author2}

@book{FORSTER1967,
title = {Chapter II - Mechanisms of Energy Transfer},
editor = {MARCEL FLORKIN and ELMER H. STOTZ},
series = {Comprehensive Biochemistry},
publisher = {Elsevier},
volume = {22},
pages = {61-80},
year = {1967},
booktitle = {Bioenergetics},
issn = {0069-8032},
doi = {https://doi.org/10.1016/B978-1-4831-9712-8.50010-2},
url = {https://www.sciencedirect.com/science/article/pii/B9781483197128500102},
author = {TH. Förster},
}

@article{Field1998,
author = {Christopher B. Field  and Michael J. Behrenfeld  and James T. Randerson  and Paul Falkowski },
title = {Primary Production of the Biosphere: Integrating Terrestrial and Oceanic Components},
journal = {Science},
volume = {281},
number = {5374},
pages = {237-240},
year = {1998},
doi = {10.1126/science.281.5374.237},
URL = {https://www.science.org/doi/abs/10.1126/science.281.5374.237},
eprint = {https://www.science.org/doi/pdf/10.1126/science.281.5374.237},
}

@article{crockford2023geologic,
  title={The geologic history of primary productivity},
  author={Crockford, Peter W and On, Yinon M Bar and Ward, Luce M and Milo, Ron and Halevy, Itay},
  journal={Current Biology},
  volume={33},
  number={21},
  pages={4741--4750},
  year={2023},
  publisher={Elsevier}
}

@article{des2000did,
  title={When did photosynthesis emerge on Earth?},
  author={Des Marais, David J},
  journal={Science},
  volume={289},
  number={5485},
  pages={1703--1705},
  year={2000},
  publisher={American Association for the Advancement of Science}
}

@article{canfield2006early,
  title={Early anaerobic metabolisms},
  author={Canfield, Don E and Rosing, Minik T and Bjerrum, Christian},
  journal={Philosophical Transactions of the Royal Society B: Biological Sciences},
  volume={361},
  number={1474},
  pages={1819--1836},
  year={2006},
  publisher={The Royal Society London}
}

@article{seager2005vegetation,
  title={Vegetation's red edge: a possible spectroscopic biosignature of extraterrestrial plants},
  author={Seager, Sara and Turner, Edwin L and Schafer, Justin and Ford, Eric B},
  journal={Astrobiology},
  volume={5},
  number={3},
  pages={372--390},
  year={2005},
  publisher={Mary Ann Liebert, Inc. 2 Madison Avenue Larchmont, NY 10538 USA}
}

@article{tucker1985african,
  title={African land-cover classification using satellite data},
  author={Tucker, Compton J and Townshend, John RG and Goff, Thomas E},
  journal={Science},
  volume={227},
  number={4685},
  pages={369--375},
  year={1985},
  publisher={American Association for the Advancement of Science}
}

@article{o1998ocean,
  title={Ocean color chlorophyll algorithms for SeaWiFS},
  author={O'Reilly, John E and Maritorena, St{\'e}phane and Mitchell, B Greg and Siegel, David A and Carder, Kendall L and Garver, Sara A and Kahru, Mati and McClain, Charles},
  journal={Journal of Geophysical Research: Oceans},
  volume={103},
  number={C11},
  pages={24937--24953},
  year={1998},
  publisher={Wiley Online Library}
}

@article{gohin2002five,
  title={A five channel chlorophyll concentration algorithm applied to SeaWiFS data processed by SeaDAS in coastal waters},
  author={Gohin, F and Druon, JN and Lampert, L},
  journal={International journal of remote sensing},
  volume={23},
  number={8},
  pages={1639--1661},
  year={2002},
  publisher={Taylor \& Francis}
}

@article{arnold2002test,
  title={A test for the search for life on extrasolar planets-Looking for the terrestrial vegetation signature in the Earthshine spectrum},
  author={Arnold, Luc and Gillet, Sophie and Lardi{\`e}re, Olivier and Riaud, Pierre and Schneider, Jean},
  journal={Astronomy \& Astrophysics},
  volume={392},
  number={1},
  pages={231--237},
  year={2002},
  publisher={EDP Sciences}
}

@article{woolf2002spectrum,
  title={The spectrum of Earthshine: a pale blue dot observed from the ground},
  author={Woolf, Neville J and Smith, Paul S and Traub, Wesley A and Jucks, Kenneth W},
  journal={The Astrophysical Journal},
  volume={574},
  number={1},
  pages={430},
  year={2002},
  publisher={IOP Publishing}
}

@article{montanes2006vegetation,
  title={Vegetation signature in the observed globally integrated spectrum of Earth considering simultaneous cloud data: applications for extrasolar planets},
  author={Montanes-Rodriguez, Pilar and Pall{\'e}, E and Goode, PR and Mart{\'\i}n-Torres, FJ},
  journal={The Astrophysical Journal},
  volume={651},
  number={1},
  pages={544},
  year={2006},
  publisher={IOP Publishing}
}

@article{montanes2005globally,
  title={Globally integrated measurements of the Earth’s visible spectral albedo},
  author={Montanes-Rodriguez, Pilar and Pall{\'e}, E and Goode, PR and Hickey, J and Koonin, SE},
  journal={The Astrophysical Journal},
  volume={629},
  number={2},
  pages={1175},
  year={2005},
  publisher={IOP Publishing}
}

@article{rodriguez2004earthshine,
  title={The earthshine spectrum},
  author={Rodriguez, P Montan{\'e}s and Pall{\'e}, E and Goode, PR and Hickey, J and Qiu, J and Yurchyshyn, Vasyl and Chu, MC and Kolbe, E and Brown, CT and Koonin, SE},
  journal={Advances in Space Research},
  volume={34},
  number={2},
  pages={293--296},
  year={2004},
  publisher={Elsevier}
}

@article{arnold2008earthshine,
  title={Earthshine observation of vegetation and implication for life detection on other planets: a review of 2001--2006 works},
  author={Arnold, Luc},
  journal={Space Science Reviews},
  volume={135},
  number={1},
  pages={323--333},
  year={2008},
  publisher={Springer}
}

@article{sterzik2012biosignatures,
  title={Biosignatures as revealed by spectropolarimetry of Earthshine},
  author={Sterzik, Michael F and Bagnulo, Stefano and Palle, Enric},
  journal={Nature},
  volume={483},
  number={7387},
  pages={64--66},
  year={2012},
  publisher={Nature Publishing Group UK London}
}

@article{sagan1993search,
  title={A search for life on Earth from the Galileo spacecraft},
  author={Sagan, Carl and Thompson, W Reid and Carlson, Robert and Gurnett, Donald and Hord, Charles},
  journal={Nature},
  volume={365},
  number={6448},
  pages={715--721},
  year={1993},
  publisher={Nature Publishing Group UK London}
}

@article{livengood2011properties,
  title={Properties of an Earth-like planet orbiting a Sun-like star: Earth observed by the EPOXI mission},
  author={Livengood, Timothy A and Deming, L Drake and A'hearn, Michael F and Charbonneau, David and Hewagama, Tilak and Lisse, Carey M and McFadden, Lucy A and Meadows, Victoria S and Robinson, Tyler D and Seager, Sara and others},
  journal={Astrobiology},
  volume={11},
  number={9},
  pages={907--930},
  year={2011},
  publisher={Mary Ann Liebert, Inc. 140 Huguenot Street, 3rd Floor New Rochelle, NY 10801 USA}
}

@article{tinetti2006detectability,
  title={Detectability of planetary characteristics in disk-averaged spectra. I: The Earth model},
  author={Tinetti, Giovanna and Meadows, Victoria S and Crisp, David and Fong, William and Fishbein, Evan and Turnbull, Margaret and Bibring, Jean-Pierre},
  journal={Astrobiology},
  volume={6},
  number={1},
  pages={34--47},
  year={2006},
  publisher={Mary Ann Liebert, Inc. 2 Madison Avenue Larchmont, NY 10538 USA}
}

@article{fujii2010colors,
  title={Colors of a second Earth: estimating the fractional areas of ocean, land, and vegetation of Earth-like exoplanets},
  author={Fujii, Yuka and Kawahara, Hajime and Suto, Yasushi and Taruya, Atsushi and Fukuda, Satoru and Nakajima, Teruyuki and Turner, Edwin L},
  journal={The Astrophysical Journal},
  volume={715},
  number={2},
  pages={866},
  year={2010},
  publisher={IOP Publishing}
}

@article{kawahara2010global,
  title={Global mapping of Earth-like exoplanets from scattered light curves},
  author={Kawahara, Hajime and Fujii, Yuka},
  journal={The Astrophysical Journal},
  volume={720},
  number={2},
  pages={1333},
  year={2010},
  publisher={IOP Publishing}
}

@article{robinson2011earth,
  title={Earth as an extrasolar planet: Earth model validation using EPOXI Earth observations},
  author={Robinson, Tyler D and Meadows, Victoria S and Crisp, David and Deming, Drake and A'hearn, Michael F and Charbonneau, David and Livengood, Timothy A and Seager, Sara and Barry, Richard K and Hearty, Thomas and others},
  journal={Astrobiology},
  volume={11},
  number={5},
  pages={393--408},
  year={2011},
  publisher={Mary Ann Liebert, Inc. 140 Huguenot Street, 3rd Floor New Rochelle, NY 10801 USA}
}

@article{fujii2012mapping,
  title={Mapping earth analogs from photometric variability: spin--orbit tomography for planets in inclined orbits},
  author={Fujii, Yuka and Kawahara, Hajime},
  journal={The Astrophysical Journal},
  volume={755},
  number={2},
  pages={101},
  year={2012},
  publisher={IOP Publishing}
}

@article{kaltenegger2007spectral,
  title={Spectral evolution of an Earth-like planet},
  author={Kaltenegger, Lisa and Traub, Wesley A and Jucks, Kenneth W},
  journal={The Astrophysical Journal},
  volume={658},
  number={1},
  pages={598},
  year={2007},
  publisher={IOP Publishing}
}

@article{wolstencroft2002photosynthesis,
  title={Photosynthesis: likelihood of occurrence and possibility of detection on Earth-like planets},
  author={Wolstencroft, RD and Raven, John A},
  journal={Icarus},
  volume={157},
  number={2},
  pages={535--548},
  year={2002},
  publisher={Elsevier}
}

@article{kiang2007spectral,
  title={Spectral signatures of photosynthesis. II. Coevolution with other stars and the atmosphere on extrasolar worlds},
  author={Kiang, Nancy Y and Segura, Ant{\'\i}gona and Tinetti, Giovanna and Govindjee and Blankenship, Robert E and Cohen, Martin and Siefert, Janet and Crisp, David and Meadows, Victoria S},
  journal={Astrobiology},
  volume={7},
  number={1},
  pages={252--274},
  year={2007},
  publisher={Mary Ann Liebert, Inc. 2 Madison Avenue Larchmont, NY 10538 USA}
}

@article{koblivzek2015ecology,
  title={Ecology of aerobic anoxygenic phototrophs in aquatic environments},
  author={Kobl{\'\i}{\v{z}}ek, Michal},
  journal={FEMS Microbiology Reviews},
  volume={39},
  number={6},
  pages={854--870},
  year={2015},
  publisher={Oxford University Press}
}

@article{bryant2007candidatus,
  title={Candidatus Chloracidobacterium thermophilum: an aerobic phototrophic acidobacterium},
  author={Bryant, Donald A and Costas, Amaya M Garcia and Maresca, Julia A and Chew, Aline Gomez Maqueo and Klatt, Christian G and Bateson, Mary M and Tallon, Luke J and Hostetler, Jessica and Nelson, William C and Heidelberg, John F and others},
  journal={Science},
  volume={317},
  number={5837},
  pages={523--526},
  year={2007},
  publisher={American Association for the Advancement of Science}
}

@article{zeng2014functional,
  title={Functional type 2 photosynthetic reaction centers found in the rare bacterial phylum Gemmatimonadetes},
  author={Zeng, Yonghui and Feng, Fuying and Medov{\'a}, Hana and Dean, Jason and Kobl{\'\i}{\v{z}}ek, Michal},
  journal={Proceedings of the National Academy of Sciences},
  volume={111},
  number={21},
  pages={7795--7800},
  year={2014},
  publisher={National Academy of Sciences}
}

@article{sanroma2013characterizing,
  title={Characterizing the purple Earth: modeling the globally integrated spectral variability of the Archean Earth},
  author={Sanrom{\'a}, E and Pall{\'e}, E and Parenteau, MN and Kiang, NY and Guti{\'e}rrez-Navarro, AM and L{\'o}pez, R and Monta{\~n}{\'e}s-Rodr{\'\i}guez, P},
  journal={The Astrophysical Journal},
  volume={780},
  number={1},
  pages={52},
  year={2013},
  publisher={IOP Publishing}
}

@article{kosmopoulos2023horizontal,
  title={Horizontal gene transfer and CRISPR targeting drive phage-bacterial host interactions and coevolution in “Pink Berry” marine microbial aggregates},
  author={Kosmopoulos, James C and Campbell, Danielle E and Whitaker, Rachel J and Wilbanks, Elizabeth G},
  journal={Applied and environmental microbiology},
  volume={89},
  number={7},
  pages={e00177--23},
  year={2023},
  publisher={American Society for Microbiology 1752 N St., NW, Washington, DC}
}

@article{coelho2024purple,
  title={Purple is the new green: biopigments and spectra of Earth-like purple worlds},
  author={Coelho, L{\'\i}gia Fonseca and Kaltenegger, Lisa and Zinder, Stephen and Philpot, William and Price, Taylor L and Hamilton, Trinity L},
  journal={Monthly Notices of the Royal Astronomical Society},
  volume={530},
  number={2},
  pages={1363--1368},
  year={2024},
  publisher={Oxford University Press}
}

@article{metz2024detectability,
  title={Detectability Simulations of a Near-infrared Surface Biosignature on Proxima Centauri b with Future Space Observatories},
  author={Metz, Connor O and Kiang, Nancy Y and Villanueva, Geronimo L and Parenteau, Mary N and Kofman, Vincent},
  journal={The Planetary Science Journal},
  volume={5},
  number={10},
  pages={228},
  year={2024},
  publisher={IOP Publishing}
}

@article{yabuzaki2017carotenoids,
  title={Carotenoids Database: structures, chemical fingerprints and distribution among organisms},
  author={Yabuzaki, Junko},
  journal={Database},
  volume={2017},
  pages={bax004},
  year={2017},
  publisher={Oxford University Press}
}

@article{vogl2011elucidation,
  title={Elucidation of the biosynthetic pathway for okenone in Thiodictyon sp. CAD16 leads to the discovery of two novel carotene ketolases},
  author={Vogl, Kajetan and Bryant, Donald A},
  journal={Journal of Biological Chemistry},
  volume={286},
  number={44},
  pages={38521--38532},
  year={2011},
  publisher={Elsevier}
}

@article{vogl2012biosynthesis,
  title={Biosynthesis of the biomarker okenone: $\chi$-ring formation},
  author={Vogl, K and Bryant, DA},
  journal={Geobiology},
  volume={10},
  number={3},
  pages={205--215},
  year={2012},
  publisher={Wiley Online Library}
}

@article{edge1997carotenoids,
  title={The carotenoids as anti-oxidants—a review},
  author={Edge, R and McGarvey, DJ and Truscott, TG},
  journal={Journal of Photochemistry and Photobiology B: Biology},
  volume={41},
  number={3},
  pages={189--200},
  year={1997},
  publisher={Elsevier}
}

@article{seel2020carotenoids,
  title={Carotenoids are used as regulators for membrane fluidity by Staphylococcus xylosus},
  author={Seel, Waldemar and Baust, Denise and Sons, Dominik and Albers, Maren and Etzbach, Lara and Fuss, Janina and Lipski, Andr{\'e}},
  journal={Scientific Reports},
  volume={10},
  number={1},
  pages={330},
  year={2020},
  publisher={Nature Publishing Group UK London}
}

@article{ma2022aromatic,
  title={Aromatic carotenoids: Biological sources and geological implications},
  author={Ma, Jian and Cui, Xingqian},
  journal={Geosystems and Geoenvironment},
  volume={1},
  number={2},
  pages={100045},
  year={2022},
  publisher={Elsevier}
}

@article{vitek2017discovery,
  title={Discovery of carotenoid red-shift in endolithic cyanobacteria from the Atacama Desert},
  author={V{\'\i}tek, Petr and Ascaso, Carmen and Artieda, Octavio and Casero, Mar{\'\i}a Cristina and Wierzchos, Jacek},
  journal={Scientific reports},
  volume={7},
  number={1},
  pages={11116},
  year={2017},
  publisher={Nature Publishing Group UK London}
}

@article{coelho2022color,
  title={Color catalogue of life in ice: Surface biosignatures on icy worlds},
  author={Coelho, L{\'\i}gia F and Madden, Jack and Kaltenegger, Lisa and Zinder, Stephen and Philpot, William and Esqu{\'\i}vel, M Gl{\'o}ria and Can{\'a}rio, Jo{\~a}o and Costa, Rodrigo and Vincent, Warwick F and Martins, Zita},
  journal={Astrobiology},
  volume={22},
  number={3},
  pages={313--321},
  year={2022},
  publisher={Mary Ann Liebert, Inc., publishers 140 Huguenot Street, 3rd Floor New~…}
}

@article{painter2001detection,
  title={Detection and quantification of snow algae with an airborne imaging spectrometer},
  author={Painter, Thomas H and Duval, Brian and Thomas, William H and Mendez, Maria and Heintzelman, Sara and Dozier, Jeff},
  journal={Applied and environmental microbiology},
  volume={67},
  number={11},
  pages={5267--5272},
  year={2001},
  publisher={American Society for Microbiology}
}

@article{dassarma2021early,
  title={Early evolution of purple retinal pigments on Earth and implications for exoplanet biosignatures},
  author={DasSarma, Shiladitya and Schwieterman, Edward W},
  journal={International Journal of Astrobiology},
  volume={20},
  number={3},
  pages={241--250},
  year={2021},
  publisher={Cambridge University Press}
}

@article{sephus2022earliest,
  title={Earliest photic zone niches probed by ancestral microbial rhodopsins},
  author={Sephus, Cathryn D and Fer, Evrim and Garcia, Amanda K and Adam, Zachary R and Schwieterman, Edward W and Kacar, Betul},
  journal={Molecular Biology and Evolution},
  volume={39},
  number={5},
  pages={msac100},
  year={2022},
  publisher={Oxford University Press}
}

@article{papageorgiou2007fast,
  title={The fast and slow kinetics of chlorophyll a fluorescence induction in plants, algae and cyanobacteria: a viewpoint},
  author={Papageorgiou, George C and Tsimilli-Michael, Merope and Stamatakis, Kostas},
  journal={Photosynthesis research},
  volume={94},
  number={2},
  pages={275--290},
  year={2007},
  publisher={Springer}
}

@article{haddock2010bioluminescence,
  title={Bioluminescence in the sea},
  author={Haddock, Steven HD and Moline, Mark A and Case, James F},
  journal={Annual review of marine science},
  volume={2},
  number={1},
  pages={443--493},
  year={2010},
  publisher={Annual Reviews}
}

@article{joiner2011first,
  title={First observations of global and seasonal terrestrial chlorophyll fluorescence from space},
  author={Joiner, J and Yoshida, Y and Vasilkov, AP and Yoshida, Y and Corp, LA and Middleton, EM},
  journal={Biogeosciences},
  volume={8},
  number={3},
  pages={637--651},
  year={2011},
  publisher={Copernicus Publications G{\"o}ttingen, Germany}
}

@article{sun2017oco,
  title={OCO-2 advances photosynthesis observation from space via solar-induced chlorophyll fluorescence},
  author={Sun, Ying and Frankenberg, Christian and Wood, Jeffery D and Schimel, David S and Jung, Martin and Guanter, Luis and Drewry, DT and Verma, Manish and Porcar-Castell, Albert and Griffis, Timothy J and others},
  journal={Science},
  volume={358},
  number={6360},
  pages={eaam5747},
  year={2017},
  publisher={American Association for the Advancement of Science}
}

@article{o2018biofluorescent,
  title={Biofluorescent worlds: Global biological fluorescence as a biosignature},
  author={O’Malley-James, Jack T and Kaltenegger, Lisa},
  journal={Monthly Notices of the Royal Astronomical Society},
  volume={481},
  number={2},
  pages={2487--2496},
  year={2018},
  publisher={Oxford University Press}
}

@article{o2018vegetation,
  title={The vegetation red edge biosignature through time on earth and exoplanets},
  author={O'Malley-James, Jack T and Kaltenegger, Lisa},
  journal={Astrobiology},
  volume={18},
  number={9},
  pages={1123--1136},
  year={2018},
  publisher={Mary Ann Liebert, Inc., publishers 140 Huguenot Street, 3rd Floor New~…}
}

@article{komatsu2023photosynthetic,
  title={Photosynthetic fluorescence from earthlike planets around sunlike and cool stars},
  author={Komatsu, Yu and Hori, Yasunori and Kuzuhara, Masayuki and Kosugi, Makiko and Takizawa, Kenji and Narita, Norio and Omiya, Masashi and Kim, Eunchul and Kusakabe, Nobuhiko and Meadows, Victoria and others},
  journal={The Astrophysical Journal},
  volume={942},
  number={2},
  pages={57},
  year={2023},
  publisher={IOP Publishing}
}

@article{kohler2021mineral,
  title={Mineral luminescence observed from space},
  author={K{\"o}hler, Philipp and Fischer, Woodward W and Rossman, George R and Grotzinger, John P and Doughty, Russell and Wang, Yujie and Yin, Yi and Frankenberg, Christian},
  journal={Geophysical Research Letters},
  volume={48},
  number={19},
  pages={e2021GL095227},
  year={2021},
  publisher={Wiley Online Library}
}

@article{miller2005detection,
  title={Detection of a bioluminescent milky sea from space},
  author={Miller, Steven D and Haddock, Steven HD and Elvidge, Christopher D and Lee, Thomas F},
  journal={Proceedings of the National Academy of Sciences},
  volume={102},
  number={40},
  pages={14181--14184},
  year={2005},
  publisher={National Academy of Sciences}
}

@article{seager2012astrophysical,
  title={An astrophysical view of Earth-based metabolic biosignature gases},
  author={Seager, Sara and Schrenk, Matthew and Bains, William},
  journal={Astrobiology},
  volume={12},
  number={1},
  pages={61--82},
  year={2012},
  publisher={Mary Ann Liebert, Inc. 140 Huguenot Street, 3rd Floor New Rochelle, NY 10801 USA}
}

@article{slaton2001estimating,
  title={Estimating near-infrared leaf reflectance from leaf structural characteristics},
  author={Slaton, Mich{\`e}le R and Raymond Hunt Jr, E and Smith, William K},
  journal={American journal of botany},
  volume={88},
  number={2},
  pages={278--284},
  year={2001},
  publisher={Wiley Online Library}
}

@article{chowdhary2019modeling,
  title={Modeling atmosphere-ocean radiative transfer: A PACE mission perspective},
  author={Chowdhary, Jacek and Zhai, Peng-Wang and Boss, Emmanuel and Dierssen, Heidi and Frouin, Robert and Ibrahim, Amir and Lee, Zhongping and Remer, Lorraine A and Twardowski, Michael and Xu, Feng and others},
  journal={Frontiers in Earth Science},
  volume={7},
  pages={100},
  year={2019},
  publisher={Frontiers Media SA}
}

@article{miller1991seasonal,
  title={Seasonal patterns in leaf reflectance red-edge characteristics},
  author={Miller, JR and Wu, Jiyou and Boyer, MG and Belanger, M and Hare, EW},
  journal={International Journal of Remote Sensing},
  volume={12},
  number={7},
  pages={1509--1523},
  year={1991},
  publisher={Taylor \& Francis}
}

@article{schwieterman2018exoplanet,
  title={Exoplanet biosignatures: a review of remotely detectable signs of life},
  author={Schwieterman, Edward W and Kiang, Nancy Y and Parenteau, Mary N and Harman, Chester E and DasSarma, Shiladitya and Fisher, Theresa M and Arney, Giada N and Hartnett, Hilairy E and Reinhard, Christopher T and Olson, Stephanie L and others},
  journal={Astrobiology},
  volume={18},
  number={6},
  pages={663--708},
  year={2018},
  publisher={Mary Ann Liebert, Inc. 140 Huguenot Street, 3rd Floor New Rochelle, NY 10801 USA}
}

@article{holland2006oxygenation,
  title={The oxygenation of the atmosphere and oceans},
  author={Holland, Heinrich D},
  journal={Philosophical Transactions of the Royal Society B: Biological Sciences},
  volume={361},
  number={1470},
  pages={903--915},
  year={2006},
  publisher={The Royal Society London}
}

@article{konhauser2017iron,
  title={Iron formations: A global record of Neoarchaean to Palaeoproterozoic environmental history},
  author={Konhauser, Kurt O and Planavsky, NJ and Hardisty, DS and Robbins, LJ and Warchola, TJ and Haugaard, R and Lalonde, SV and Partin, CA and Oonk, PBH and Tsikos, H and others},
  journal={Earth-Science Reviews},
  volume={172},
  pages={140--177},
  year={2017},
  publisher={Elsevier}
}

@article{allwood2006stromatolite,
  title={Stromatolite reef from the Early Archaean era of Australia},
  author={Allwood, Abigail C and Walter, Malcolm R and Kamber, Balz S and Marshall, Craig P and Burch, Ian W},
  journal={Nature},
  volume={441},
  number={7094},
  pages={714--718},
  year={2006},
  publisher={Nature Publishing Group UK London}
}

@article{planavsky2014evidence,
  title={Evidence for oxygenic photosynthesis half a billion years before the Great Oxidation Event},
  author={Planavsky, Noah J and Asael, Dan and Hofmann, Axel and Reinhard, Christopher T and Lalonde, Stefan V and Knudsen, Andrew and Wang, Xiangli and Ossa Ossa, Frantz and Pecoits, Ernesto and Smith, Albertus JB and others},
  journal={Nature Geoscience},
  volume={7},
  number={4},
  pages={283--286},
  year={2014},
  publisher={Nature Publishing Group UK London}
}

@article{matsuo2025archaean,
  title={Archaean green-light environments drove the evolution of cyanobacteria’s light-harvesting system},
  author={Matsuo, Taro and Ito-Miwa, Kumiko and Hoshino, Yosuke and Fujii, Yuri I and Kanno, Satomi and Fujimoto, Kazuhiro J and Tsuji, Rio and Takeda, Shinnosuke and Onami, Chieko and Arai, Chihiro and others},
  journal={Nature Ecology \& Evolution},
  pages={1--14},
  year={2025},
  publisher={Nature Publishing Group UK London}
}

@article{kiyokawa2021hydrothermal,
  title={Hydrothermal formation of iron-oxyhydroxide chimney mounds in a shallow semi-enclosed bay at Satsuma Iwo-Jima Island, Kagoshima, Japan},
  author={Kiyokawa, Shoichi and Kuratomi, Takashi and Hoshino, Tatsuhiko and Goto, Shusaku and Ikehara, Minoru},
  journal={Bulletin},
  volume={133},
  number={9-10},
  pages={1890--1908},
  year={2021},
  publisher={Geological Society of America}
}

@article{konhauser2007decoupling,
  title={Decoupling photochemical Fe (II) oxidation from shallow-water BIF deposition},
  author={Konhauser, Kurt O and Amskold, Larry and Lalonde, Stefan V and Posth, Nicole R and Kappler, Andreas and Anbar, Ariel},
  journal={Earth and Planetary Science Letters},
  volume={258},
  number={1-2},
  pages={87--100},
  year={2007},
  publisher={Elsevier}
}

@article{lancaster2018decadal,
  title={The Decadal Survey in Astronomy and Astrophysics 2020 (Astro 2020)},
  author={Lancaster, James C},
  journal={NSF Award Number 1852611. Directorate for Mathematical and Physical Sciences},
  volume={18},
  number={1852611},
  pages={52611},
  year={2018}
}

\end{document}